\documentclass[aps,pre,showpacs,floatfix,longbibliography]{revtex4-1}
\usepackage{amsmath,amssymb,color,graphicx,array,bm,url}

\begin{document}

\title{Convection in the active layer speeds up permafrost thaw
in coarse grained soils}

\author{M. Magnani$^{1,2,3}$}
\author{S. Musacchio$^{3,4}$}
\author{A. Provenzale$^{1,2}$}
\author{G. Boffetta$^{3,4}$}
\thanks{Corresponding author}
\email{marta.magnani@unito.it}
\affiliation{$^1$Institute of Geoscience and Earth Resources, National Research Council (CNR-IGG), Via Valperga Caluso 35, 10125 Torino, Italy}
\affiliation{$^2$National Biodiversity Future Center, Palermo 90133, Italy}
\affiliation{$^3$INFN, Sezione di Torino, via P. Giuria 1, 10125 Torino, Italy.}
\affiliation{$^4$Dipartimento di Fisica, Universit\`a degli Studi di Torino, via P. Giuria 1, 10125 Torino, Italy.}


\begin{abstract}
Permafrost thaw is a major concern raised by the ongoing climate change. An
understudied phenomenon possibly affecting the pace of permafrost thaw is the
onset of convective motions within the active layer caused by the density
anomaly of water.  
Here, we explore the effects of groundwater convection on
permafrost thawing using a model that accounts for ice - water phase
transitions, coupled with the dynamics of the temperature field transported by
the Darcy's flow across a porous matrix.  
Numerical simulations of this model show that ice thawing in
the presence of convection is much faster than in the diffusive case
and deepens at a constant velocity proportional to the soil permeability.
A scaling argument is able to predict correctly the
asymptotic velocity. Since in the convective regime the heat transport
is mediated by the coherent motion of thermal plumes across the thawed layer,
we find that the depth of the thawing interface becomes highly heterogeneous.
\end{abstract}

\date{\today}

\maketitle
{\it Introduction.} 
The thawing of permafrost - ground that has been frozen for at least two
consecutive years - poses important societal and climatic challenges owing to
its control on soil stability, health hazards and mobilization of compounds
trapped in cold soils \cite{hjort2022,karjalainen2019,hjort2018,colombo2018}. 
Permafrost evolution is deeply tied to the thermal and hydrological dynamics at
ground surface and within the so-called ``active layer", that is the surface
layer subject to annual thawing and freezing usually present above permafrost.
The dynamics of both the permafrost and the active layer are mostly estimated
on the basis of models \cite{chadburn2017,obu2019}, while direct measurements
are rather sparse and spotty \cite{karjalainen2019,obu2019}.  

\begin{figure}[h!]
\raggedright \hspace{95pt} \textbf{(a)} \hspace{235pt} \textbf{(b)} \\
\includegraphics[width=0.4\columnwidth]{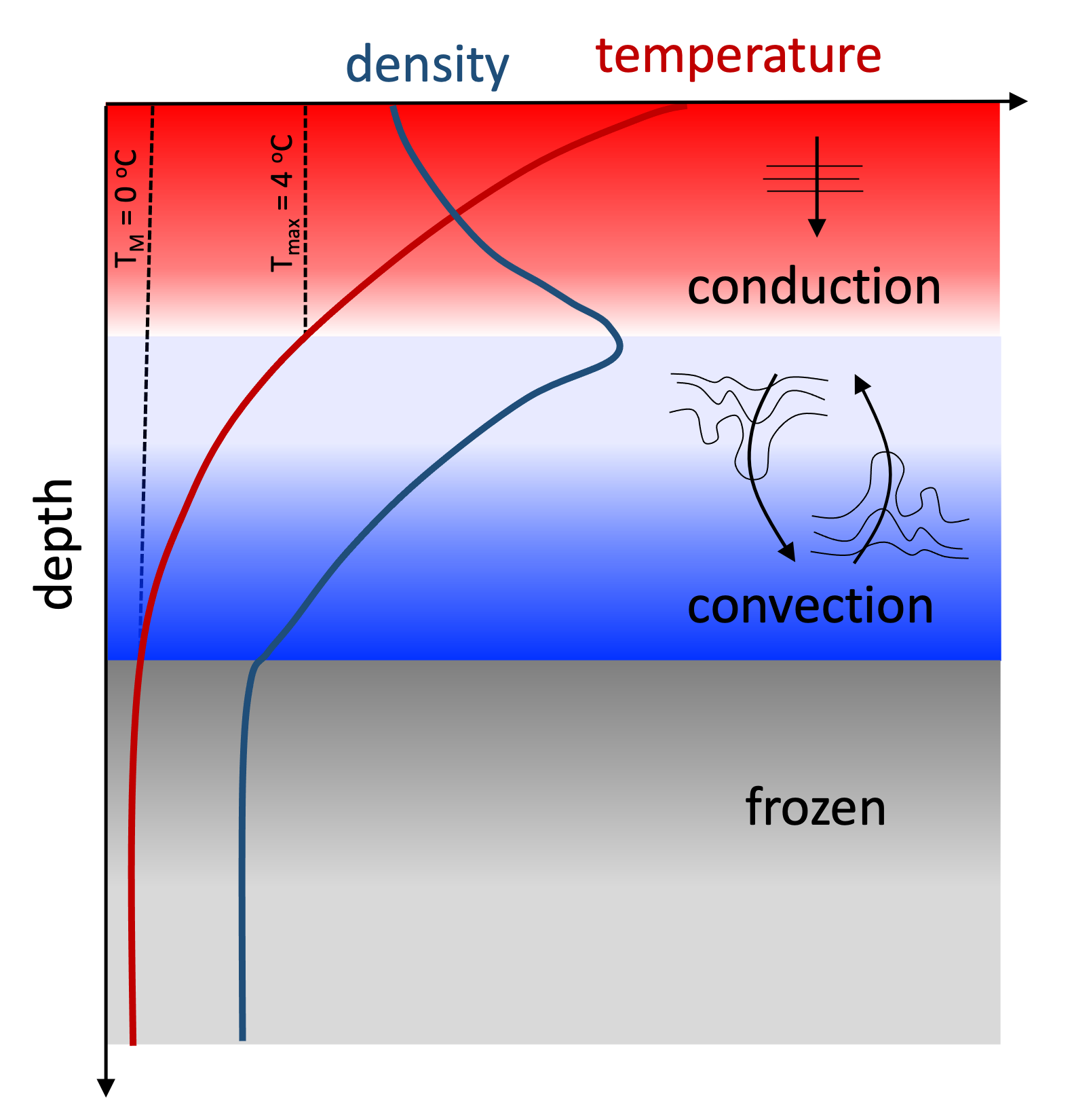}
\includegraphics[width=0.59\columnwidth]{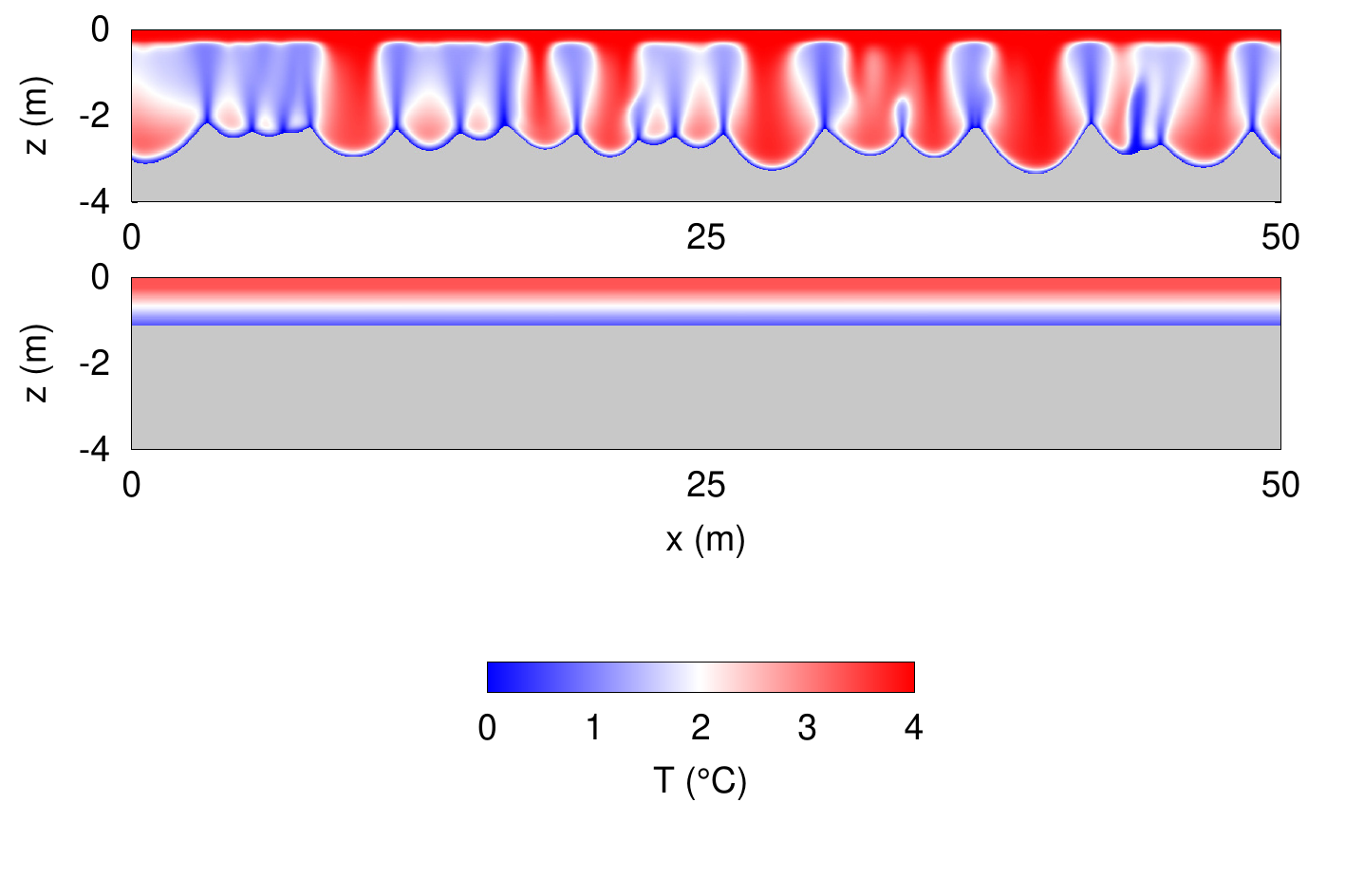}
\caption{(a) Schematic dynamics of the active layer considered in this work.
In the top layer, where $T>4 \, ^o C$, the fluid stratification is stable, 
resulting in a diffusive heat transfer. Below this layer the unstable 
stratification can produce a convective
layer characterized by the presence of thermal plumes. 
The permafrost layer where $T<0 \, ^o C$ is represented in grey. 
Conceptual representations of temperature and density 
water profiles are sketched by the red and blue line, respectively.
(b) Vertical sections of the temperature fields for run
C (upper panel) and D (lower panel) at time $t=150$~days, starting
from the same initial conditions.
Colors indicate the temperature field. The grey area
represents the frozen region where $\phi=1$.}
\label{fig1}
\end{figure}

Climate models mostly project permafrost evolution accounting for
the vertical conductive heat transfer across the active layer
\cite{dankers2011,wania2009,guimberteau2018,lawrence2008,andresen2020}. 
Nonetheless, a crucial role in determining the speed of permafrost thaw 
can be played by fluid convection in the active layer soil matrix
at large permeability,
sustained by the density anomaly of water. During
the thawing process, the temperature within the active layer decreases with
depth from the surface value to the freezing temperature met at the contact with the permafrost. 
Across this thermal profile, fluid density varies nonlinearly (Fig.~\ref{fig1}a): below $T \simeq 4~^oC$ water
has a negative thermal expansion coefficient, i.e., its density increases with
temperature, whereas above this value the temperature-density relationship is inverted. 
Therefore, in the lower part of the active layer 
where $0~^oC < T < 4~^oC$, soil water is unstably stratified and convection 
can occur.

In this work we study the 
effect of active layer fluid motion on permafrost thaw. We consider
an homogeneous-isotropic porous medium saturated with
water in the presence of gravity. For the application to permafrost thaw, 
we assume that the lower part of the domain is frozen (at temperature
$T_0<T_M=0 \, ^oC$) while an upper thin porous layer 
is already melted and 
in contact to the surface at a temperature $T_1>T_M$ 
(see the scheme in Fig.~\ref{fig1}a).
Three-dimensional direct numerical simulations of this system show that, 
depending on the permeability of the porous medium,
convective motions can develop in the active layer. This accelerates
substantially the thawing process with respect to the diffusive behavior
and produces a complex water-ice interface which deepens ballistically
in time.

{\it Mathematical model and numerical simulations.}
The physics of thawing in a porous medium is described by a phase field 
method, a convenient numerical tool for 
simulating the dynamics of multiphase fluids in the presence of 
phase transitions 
\cite{beckermann1999modeling,favier2019rayleigh,yang2022abrupt}.
The physical state of water is represented by a continuous phase field
$\phi({\bm x},t)$ which takes the value $\phi=0$ in the liquid phase 
and $\phi=1$ in the solid (ice) phase. The equation for the phase
field is a Allen-Cahn type equation \cite{hester2020improved} coupled with 
an advection-diffusion equation for the temperature
field $T({\bf x},t)$ and the Darcy's equation, which describes the 
velocity field ${\bf u}({\bf x},t)$ in the porous medium 
\begin{equation}
{\bm u} = {k \over \mu \varphi} (-{\bm \nabla}p+(1-\phi)^2 {\bm g} \rho)
\label{eq1}
\end{equation}
\begin{equation}
{\partial T \over \partial t} + {\bm u} \cdot {\bm \nabla} T =
\kappa \nabla^2 T + \Delta T St
{\partial \phi \over \partial t}
\label{eq2}
\end{equation}
\begin{equation}
{\partial \phi \over \partial t} =
{6 \kappa' \over 5 St} \left[ \nabla^2 \phi -
{1 \over \delta^2} \phi (1-\phi)\left(1 - 2 \phi + 
{T - T_M \over \Delta T}\right) \right]
\label{eq3}
\end{equation}
where ${\bf g}=(0,0,-g)$ represents gravity, $k$ is the permeability,
$\mu$ is the fluid viscosity, $\varphi$ the porosity, $p$ is the 
total pressure and $\kappa$ represents an effective thermal diffusivity 
(linear combination 
of the solid and the liquid contributions \cite{rees2000vertical}).
$St=L/(c_p \Delta T)$ is the dimensionless Stefan number 
defined in terms of the latent heat for unit mass $L$, the specific 
heat capacity of water $c_p$ and the temperature jump $\Delta T=T_1-T_0$,
while $\delta$ and $\kappa'$ 
represent the interface thickness and mobility.
Finally, density and temperature of water are related by the
empirical non-monotonic model, valid around the temperature 
$T_{max}=3.98 \, ^oC$ of maximal density, 
$\rho=\rho_0 (1 - \alpha^* \lvert T - T_{max} \rvert ^q )$ 
where $\alpha^*$ is a generalized thermal expansion coefficient,
$q =1.895$ and $\rho_0 = 999.97  \, kg/m^3$ \cite{wang2021growth}.
We remark that the model (\ref{eq1}-\ref{eq3}) assumes 
local thermal equilibrium by which a single 
temperature field describes the solid and the fluid phases
in the porous medium \cite{lapwood1948convection}. 
More complex, non-equilibrium, models 
which take into account heat exchanges within the 
solid matrix have been proposed
\cite{rees2000vertical,banu2002onset} and could be used for future studies in 
the present setup.

Because of the density anomaly, the thawed water has an unstable
density stratification in the layer where $T_M < T < T_{max}$. 
The dimensionless number which controls the dynamics of this fluid layer
is the Rayleigh-Darcy number
\begin{equation}
Ra = {g \Delta \rho k H \over \varphi \mu \kappa} 
\label{eq4}
\end{equation}
where $H$ and $\Delta \rho$ are respectively the thickness of and
the density difference in the convective layer.
For values of $Ra$ below the critical value $Ra_c=4 \pi^2$, 
the heat transfer across the active layer is diffusive and the 
fluid is at rest \cite{nield2006convection}.
In contrast, when $Ra>Ra_c$ the unstable density stratification
causes a convective motion which displaces the fluid and the temperature field.

We performed direct numerical simulations of equations
(\ref{eq1}-\ref{eq3}) by a fully parallel (MPI) pseudo-spectral
code up to resolution $2048 \times 2048 \times 512$.
Boundary conditions on the top and bottom layers
are imposed by penalization terms, while we use periodic condition
in the horizontal directions. 
In all the simulations the fluid is initially at rest
and the vertical temperature profile within the active layer 
of depth $H_0$ is a linear ramp,
connecting the temperature $T_0=-1 ^o C$ in the permafrost bulk to 
$T_1=T_{max}$ at the ground surface. 
In order to destabilize the flow, a small random 
perturbation is added to the initial fluid temperature profile. 
Units for space and time are meters and days respectively and
for the physical parameters we set $\kappa=0.012 \, m^2/day$,
$\mu=86.0 \, kg/(m \, day)$, $L=3.3 \times 10^{5} \, J/kg$ 
(for simplicity we use the same values for liquid and ice phase). 
Three different runs have been performed, all with porosity $\varphi=0.5$,
with permeabilities $k=4.4 \times 10^{-9} m^2$ (A), $k=5.9 \times 10^{-9} m^2$ 
(B) and $k=8.8 \times 10^{-9} m^2$ (C), typical of water in gravel. 
As a benchmark, we also consider an additional run with much smaller 
permeability $k=6 \times 10^{-11} m^2$ which remains in the diffusive regime.
The numerical method has been tested in the case of a purely 
diffusive process, i.e. with ${\bm u}=0$ in (\ref{eq2}), 
corresponding to the standard Stefan problem, for which analytical predictions
are available.

{\it Results.}
Assuming an initial small thickness $H_0$,
such that $Ra<Ra_c$, the early stage of the thawing process is characterized by
diffusive heat flux from the surface.
Consequently, the width of the liquid layer
is expected to increase as 
$H(t) \propto t^{1/2}$ \cite{rubinshteuin1971stefan}.
This growth is accompanied by the increase of Rayleigh-Darcy number 
(\ref{eq4}), which eventually becomes larger than $Ra_c$ causing the 
onset of convection.
It is well known that, in general, convection is much more efficient to
transfer heat than diffusion, and porous convection is not an 
exception \cite{hewitt2020,boffetta2020}.
Therefore, in the convective regime $H(t)$ is expected to grow much 
faster than in the diffusive case. 
This results in a further increase of $Ra$ and of the convective motion, 
which accelerates the thawing process.


\begin{figure}[h!]
\begin{center}
\raggedright \hspace{100pt} \textbf{(a)} \hspace{223pt} \textbf{(b)} \\
\includegraphics[width=0.4\textwidth]{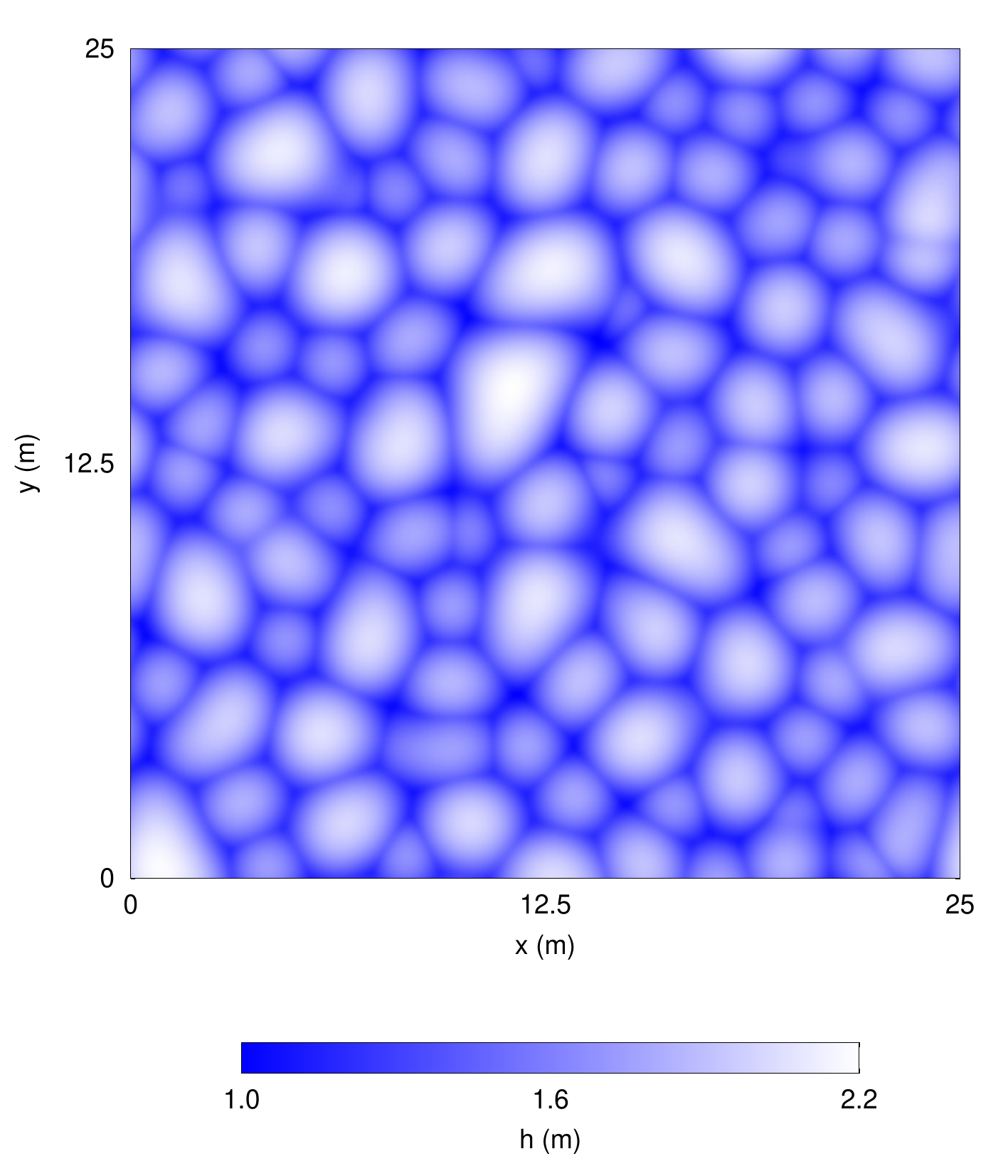}
\includegraphics[trim=0 -2cm 0 0,width=0.54\textwidth]{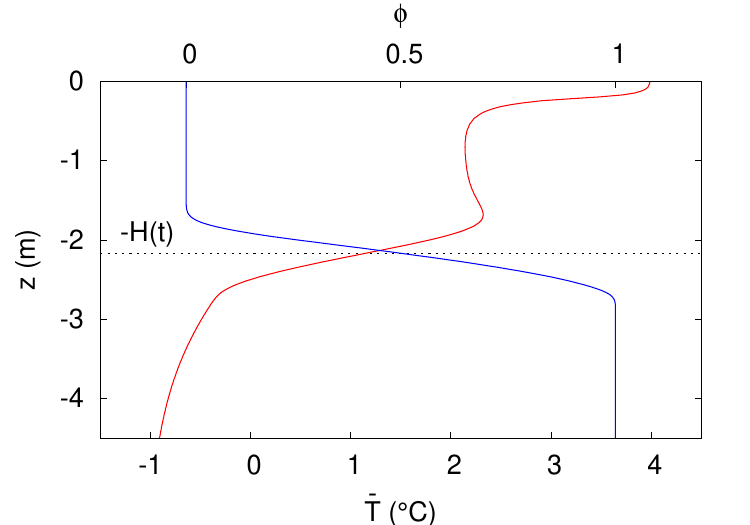}
\end{center}
\caption{(a) Thawing surface depth $h(x,y)$, defined from the phase 
field as the points where $\phi(x,y,h)=1/2$, for run $C$ 
at time $t=150 \, days$. The average position of the surface 
is $H=1.6 \, m$.
(b) Vertical temperature profile $\overline{T}(z)$ (red line) and
phase field profile $\overline{\phi}(z)$ (blue line)
at time $t=150 \, days$.
The black dotted line represents the average position of the interface
defined as $\overline{\phi}(-H)=1/2$.
Results from run $C$.}
\label{fig2}
\end{figure}

A first insight into the difference between the conductive and convective
dynamics can be obtained from the active layer temperature field. 
Figure~\ref{fig1}b shows two vertical sections on the $(x,z)$ plane of the water temperature
field (colors) taken at $t=150 \, days$ for the conductive case (run D, lower
panel),
and one convective case (run C, upper panel). 
The frozen part is represented in grey.

When the permeability is small (run D) the heat transfer in the active layer
is purely diffusive, resulting in a uniform temperature gradient in the liquid
layer and a flat fluid-ice interface.  
Conversely, the convective cases (such as run C in Fig.~\ref{fig1}b) are 
characterized by the presence of thermal plumes that are responsible for 
the efficient exchange of heath across the active layer and speed up 
the thawing process. 
Thermal plumes are generated close to the layer at temperature $T_{max}$
and they move downwards causing an inhomogeneous heating and thawing 
of the ice. As a consequence, the fluid-ice interface is not flat, 
and ice and water coexist at a 
given depth $z$, in clear contrast with the diffusive case 
(compare the top and bottom simulation in Fig.~\ref{fig1}b).

Thermal plumes that have reached different depths 
in the thawing process create complex patterns 
at the {\it thawing surface}
$h(x,y,t)$, defined in this study as the depth where 
the local phase field $\phi({\bm x},t)=1/2$.
One example of this surface is shown 
in Fig.~\ref{fig2}a for run $C$ at time $t=150$ days. 
The average depth reached by convection is in this case
$H(t) \simeq 1.6 \, m$, but we observe local fluctuations 
ranging from $1 \, m$ up to $2.2 \, m$ below downwelling plumes.
Remarkably, similar patterns have been recently reported as
the results of dissolution processes observed both in geological sites
and in laboratory experiments in bulk flows 
\cite{chaigne2023emergence}.

In order to compare quantitatively the conductive and convective 
processes, we compute the vertical profiles of temperature field
$\overline{T}(z)$ and of phase field $\overline{\phi}(z)$, where
the overbar represents the average over horizontal planes. 
Figure~\ref{fig2}b shows one example of these profiles for the 
run C. In the temperature profile we recognize the presence of a
boundary layer close to the upper surface at temperature $T_1$, 
followed by a non-monotonic temperature pattern, due to the 
presence of thermal plumes which produce a local temperature
maximum. 
Below this maximum, the temperature decreases sharply, while the phase field
profile $\overline{\phi}(z)$ increases. The region in which 
$0<\overline{\phi}<1$ corresponds to the thawing surface with 
the coexistence, at a given depth, of zones which are
liquid ($\phi=0$) with others still frozen ($\phi=1$). 
The depth at which $\overline{\phi}(z)=1/2$ is used to define the 
average vertical position $H(t)$ of the water-ice interface (dotted line 
in Fig.~\ref{fig2}). 

\begin{figure}[h!]
\raggedright \hspace{130pt} \textbf{(a)} \hspace{230pt} \textbf{(b)} \\
\includegraphics[width=0.49\textwidth]{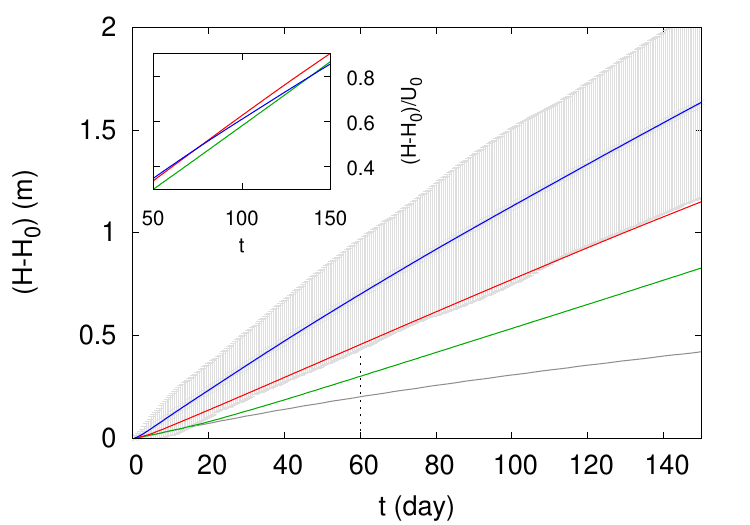}
\includegraphics[width=0.49\textwidth]{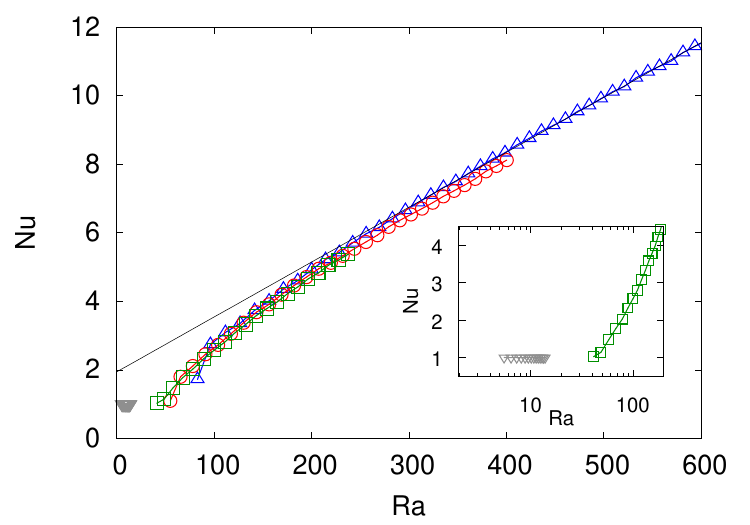}
\caption{(a) Average displacement of the water-ice interface as a function of time 
for the three runs in the convective regime A (green line), 
B (red line) and C (blue line) and for run D
in the diffusive regime (grey line). $H_0=0.33 \, m$ is the 
initial position of the interface at $t=0$.
The grey area represents the width of the
thawing layer ($0<\overline{\phi}(z)<1$) for run C. The 
vertical dashed line is a guide 
for the $t=60$ days.
Inset: displacement rescaled with the characteristic velocity $U_0$
in the last part of the evolution.
(b) Nusselt number $Nu$ as a function of the Rayleigh-Darcy
number $Ra$ for run $A$ (green squares), $B$ (red circles),
$C$ (blue triangles) and $D$ (grey inverted triangles). 
The black line represent the linear fit $Nu=0.016 Ra+const$.
Inset: The same data in semi-logarithmic scales for the lower range
of $Ra$.}
\label{fig3}
\end{figure}

The displacement of the interface $H(t)$ as a function of time
is shown in Fig.~\ref{fig3}a. 
The convective case with lowest permeability (run A, green line) is initially
below the critical Rayleigh-Darcy number and therefore in a first phase 
it follows the diffusive Stefan solution (grey line). 
At about $t=15 \, days$ convection sets in and
the displacement of the interface accelerates with respect to the 
diffusive case. For larger values of the permeability
(run B, red line and run C, blue line)
the convective motion develops since the beginning.
The effect of convection on the thawing process
is evident: after $60$ days (vertical dashed line) 
the interface has displaced, in the case of 
the highest permeability, of about $0.7 \, m$ from the starting position, 
that is more than three times what is predicted in the absence 
of convection (grey line).

At long times, the convection-driven motion of the interface attains a linear
law
$H \propto t$, instead of the diffusive growth predicted by the Stefan 
model. This can be understood from the mathematical model governing
the process. From the Darcy equation (\ref{eq1}) one can define 
a characteristic velocity $U_0 = \Delta \rho k g/(\mu \varphi)$ 
which represents the typical speed of thermal plumes. 
By rescaling the different trajectories of $H(t)$ in Fig.~\ref{fig3}a 
with $U_0$ one obtains the collapse on the linear behavior
$H(t)=\alpha U_0 t$ with $\alpha \simeq 0.0055$,
as shown in the inset of Fig.~\ref{fig3}a. 
Hence, by changing the permeability value (or the other
parameters in $U_0$) one obtains different velocities of the thawing 
interface in the convective regime.
We remark that the dimensionless coefficient
$\alpha$ is here much smaller than the one observed in 
Rayleigh-Taylor convection in porous media \cite{boffetta2020}
since a significant fraction of the energy is used to thaw the
ice phase.

Together with $H$, we find that also the extension of the 
layer where $0<\overline{\phi}(z)<1$ grows with time, 
as shown by the grey area in Fig.~\ref{fig3}a for run C.
This region corresponds to the coexistence of zones 
that are already liquid ($\phi=0$) 
with others still frozen ($\phi=1$) at a given depth,
i.e. the vertical extension of the thawing surface shown in Fig.~\ref{fig2}a.


The enhancement of the heat transport due to the convection
with respect to the thermal diffusion
is quantified by the Nusselt number $Nu$,
defined as the ratio of the total (convective plus
diffusive) heat flux to the diffusive one:
\begin{equation}
Nu = {\langle u_z T \rangle H \over \kappa (T_{max}-T_M)} + 1
\label{eq5}
\end{equation}
where $\langle \cdots \rangle$ indicates the average over
the convective region of depth $H$ and $u_z$ is the vertical
velocity. 
Figure~\ref{fig3}b shows the evolution of the Nusselt number 
(\ref{eq5}) as a function of the Rayleigh-Darcy number (\ref{eq4})
for the four runs. 
We recall that since the thickness of the active
layer $H$ grows with time, so does the Rayleigh-Darcy number and 
therefore each simulation spans a range of $Ra$.
Given the small density differences of water between $T_M$ and 
$T_{max}$, the Rayleigh-Darcy number remains moderate
for all the permeability values investigated here.
As shown in the inset of Figure~\ref{fig3}b, the Nusselt number 
in the run D at the lowest permeability is always $Nu=1$, because it 
remains in the diffusive regime ($Ra<Ra_c$).
For the other three simulations we see that, for sufficiently
large values of the Rayleigh-Darcy number, 
$Nu$ increases with $Ra$ following a linear law. 
This can be understood by a scaling argument based on the 
definitions of the two dimensionless numbers. 
Indeed, from (\ref{eq4}) we see that,
at fixed values of the fluid parameters, $Ra$ is proportional to
the thickness $H$ of the fluid layer. From Eq. (\ref{eq5}), assuming that
the correlation between the vertical velocity and the temperature field
can be simply expressed as $\langle u_z T \rangle = \beta U_0 \Delta T$
(where $\beta$ is a dimensionless constant) we have the 
prediction of a linear scaling $Nu = \beta Ra$, as indeed observed in 
Fig.~\ref{fig3}b, with $\beta \simeq 0.016$.
We remark that, in spite of the different thawing speed for
the different permeability values, 
Fig.~\ref{fig3}b indicates a universal $Ra-Nu$ relation, almost independent
of the permeability $k$. This suggests that the correlation between the vertical
velocity and the temperature, which enters in Eq. (\ref{eq5}), 
is not affected by the permeability of the porous medium.

To conclude, we have studied how convection 
accelerates the pace of permafrost thawing compared to the
conduction process in gravel-dominated soils. 
With the onset of convection, the
ice-water interface deepens linearly in time, in clear
contrast with the diffusive law expected from the conductive case.
Another consequence of convection is the formation of a complex thawing
interface, charaterized by spatial patterns, that can be used as a diagnostic for
the presence of convection in the active layer. 
Remarkably, similar patterns have been observe in dissolution
processes in bulk flows \cite{chaigne2023emergence}. 
Further investigations are needed to clarify 
whether the observed thawing surface can be explained within 
the same theoretical framework. 

Convective flows have been already mentioned in Arctic field studies
\cite{boike1998,weismuller2011,hollesen2011}, 
thus confirming the possibility
of this process in natural conditions. 
According to the present study, 
the onset of convection is mostly controlled by the soil permeability. 
Indeed in saturated soils, large permeability values are needed to
reach the condition for the fluid density stratification to become 
unstable and undergo convective motions. 
Using typical water parameters, such a condition is met for instance in
rocky and sandy-gravel soils \cite{gleeson2011,chapuis2012,ebel2019}. 
Fire events in arctic and
subarctic ecosystems have also been suggested to increase the connectivity of soil pores,  i.e. soil permeability, thus allowing for advective fluid flows that enhance post-fire permafrost
degradation \cite{zipper2018,ebel2019}. 
Instead, if the active layer is characterized by a soil matrix with low permeability
the onset of convection is usually prevented.

Despite using a simplified representation of the active layer system, 
where for instance soil characteristics are kept constant in time and across soil depths,  
we showed the importance of considering convective processes in
cold soils as their effects can already manifest on short time 
scales of the order of one month. Hence, the effects of convection
can already be felt within one summer period, when the ground surface
temperature can be much higher than $4~^oC$. 
These results suggest that climate models accounting for only heat conduction
may underestimate the permafrost thawing depth in most permeable soils, 
where soil water convection in the active layer can occur. 
In these cases, the framework proposed here may serve 
to obtain useful parameterisations to be implemented in climate models.  

%
%

{\it Acknowledgements.} 
We acknowledge HPC CINECA for computing resources (INFN-CINECA Grant 
INFN23-FieldTurb).
M.M., and A.P. acknowledge the Italian National Biodiversity Future Center
(NBFC) National Recovery and Resilience Plan (NRRP; mission 4, component 2,
investment 1.4 of the Ministry of University and Research, funded by the
European Union -  NextGenerationEU; project code CN00000033).

\bibliography{biblio}

\begin{thebibliography}{31}%
\makeatletter
\providecommand \@ifxundefined [1]{%
 \@ifx{#1\undefined}
}%
\providecommand \@ifnum [1]{%
 \ifnum #1\expandafter \@firstoftwo
 \else \expandafter \@secondoftwo
 \fi
}%
\providecommand \@ifx [1]{%
 \ifx #1\expandafter \@firstoftwo
 \else \expandafter \@secondoftwo
 \fi
}%
\providecommand \natexlab [1]{#1}%
\providecommand \enquote  [1]{``#1''}%
\providecommand \bibnamefont  [1]{#1}%
\providecommand \bibfnamefont [1]{#1}%
\providecommand \citenamefont [1]{#1}%
\providecommand \href@noop [0]{\@secondoftwo}%
\providecommand \href [0]{\begingroup \@sanitize@url \@href}%
\providecommand \@href[1]{\@@startlink{#1}\@@href}%
\providecommand \@@href[1]{\endgroup#1\@@endlink}%
\providecommand \@sanitize@url [0]{\catcode `\\12\catcode `\$12\catcode
  `\&12\catcode `\#12\catcode `\^12\catcode `\_12\catcode `\%12\relax}%
\providecommand \@@startlink[1]{}%
\providecommand \@@endlink[0]{}%
\providecommand \url  [0]{\begingroup\@sanitize@url \@url }%
\providecommand \@url [1]{\endgroup\@href {#1}{\urlprefix }}%
\providecommand \urlprefix  [0]{URL }%
\providecommand \Eprint [0]{\href }%
\providecommand \doibase [0]{http://dx.doi.org/}%
\providecommand \selectlanguage [0]{\@gobble}%
\providecommand \bibinfo  [0]{\@secondoftwo}%
\providecommand \bibfield  [0]{\@secondoftwo}%
\providecommand \translation [1]{[#1]}%
\providecommand \BibitemOpen [0]{}%
\providecommand \bibitemStop [0]{}%
\providecommand \bibitemNoStop [0]{.\EOS\space}%
\providecommand \EOS [0]{\spacefactor3000\relax}%
\providecommand \BibitemShut  [1]{\csname bibitem#1\endcsname}%
\let\auto@bib@innerbib\@empty
\bibitem [{\citenamefont {Hjort}\ \emph {et~al.}(2022)\citenamefont {Hjort},
  \citenamefont {Streletskiy}, \citenamefont {Dor{\'e}}, \citenamefont {Wu},
  \citenamefont {Bjella},\ and\ \citenamefont {Luoto}}]{hjort2022}%
  \BibitemOpen
  \bibfield  {author} {\bibinfo {author} {\bibfnamefont {Jan}\ \bibnamefont
  {Hjort}}, \bibinfo {author} {\bibfnamefont {Dmitry}\ \bibnamefont
  {Streletskiy}}, \bibinfo {author} {\bibfnamefont {Guy}\ \bibnamefont
  {Dor{\'e}}}, \bibinfo {author} {\bibfnamefont {Qingbai}\ \bibnamefont {Wu}},
  \bibinfo {author} {\bibfnamefont {Kevin}\ \bibnamefont {Bjella}}, \ and\
  \bibinfo {author} {\bibfnamefont {Miska}\ \bibnamefont {Luoto}},\ }\bibfield
  {title} {\enquote {\bibinfo {title} {Impacts of permafrost degradation on
  infrastructure},}\ }\href@noop {} {\bibfield  {journal} {\bibinfo  {journal}
  {Nat. Rev. Earth Environ.}\ }\textbf {\bibinfo {volume} {3}},\ \bibinfo
  {pages} {24--38} (\bibinfo {year} {2022})}\BibitemShut {NoStop}%
\bibitem [{\citenamefont {Karjalainen}\ \emph {et~al.}(2019)\citenamefont
  {Karjalainen}, \citenamefont {Aalto}, \citenamefont {Luoto}, \citenamefont
  {Westermann}, \citenamefont {Romanovsky}, \citenamefont {Nelson},
  \citenamefont {Etzelm{\"u}ller},\ and\ \citenamefont
  {Hjort}}]{karjalainen2019}%
  \BibitemOpen
  \bibfield  {author} {\bibinfo {author} {\bibfnamefont {Olli}\ \bibnamefont
  {Karjalainen}}, \bibinfo {author} {\bibfnamefont {Juha}\ \bibnamefont
  {Aalto}}, \bibinfo {author} {\bibfnamefont {Miska}\ \bibnamefont {Luoto}},
  \bibinfo {author} {\bibfnamefont {Sebastian}\ \bibnamefont {Westermann}},
  \bibinfo {author} {\bibfnamefont {Vladimir~E}\ \bibnamefont {Romanovsky}},
  \bibinfo {author} {\bibfnamefont {Frederick~E}\ \bibnamefont {Nelson}},
  \bibinfo {author} {\bibfnamefont {Bernd}\ \bibnamefont {Etzelm{\"u}ller}}, \
  and\ \bibinfo {author} {\bibfnamefont {Jan}\ \bibnamefont {Hjort}},\
  }\bibfield  {title} {\enquote {\bibinfo {title} {Circumpolar permafrost maps
  and geohazard indices for near-future infrastructure risk assessments},}\
  }\href@noop {} {\bibfield  {journal} {\bibinfo  {journal} {Sci. Data}\
  }\textbf {\bibinfo {volume} {6}},\ \bibinfo {pages} {1--16} (\bibinfo {year}
  {2019})}\BibitemShut {NoStop}%
\bibitem [{\citenamefont {Hjort}\ \emph {et~al.}(2018)\citenamefont {Hjort},
  \citenamefont {Karjalainen}, \citenamefont {Aalto}, \citenamefont
  {Westermann}, \citenamefont {Romanovsky}, \citenamefont {Nelson},
  \citenamefont {Etzelm{\"u}ller},\ and\ \citenamefont {Luoto}}]{hjort2018}%
  \BibitemOpen
  \bibfield  {author} {\bibinfo {author} {\bibfnamefont {Jan}\ \bibnamefont
  {Hjort}}, \bibinfo {author} {\bibfnamefont {Olli}\ \bibnamefont
  {Karjalainen}}, \bibinfo {author} {\bibfnamefont {Juha}\ \bibnamefont
  {Aalto}}, \bibinfo {author} {\bibfnamefont {Sebastian}\ \bibnamefont
  {Westermann}}, \bibinfo {author} {\bibfnamefont {Vladimir~E}\ \bibnamefont
  {Romanovsky}}, \bibinfo {author} {\bibfnamefont {Frederick~E}\ \bibnamefont
  {Nelson}}, \bibinfo {author} {\bibfnamefont {Bernd}\ \bibnamefont
  {Etzelm{\"u}ller}}, \ and\ \bibinfo {author} {\bibfnamefont {Miska}\
  \bibnamefont {Luoto}},\ }\bibfield  {title} {\enquote {\bibinfo {title}
  {{Degrading permafrost puts Arctic infrastructure at risk by mid-century}},}\
  }\href@noop {} {\bibfield  {journal} {\bibinfo  {journal} {Nat. Commun.}\
  }\textbf {\bibinfo {volume} {9}},\ \bibinfo {pages} {5147} (\bibinfo {year}
  {2018})}\BibitemShut {NoStop}%
\bibitem [{\citenamefont {Colombo}\ \emph {et~al.}(2018)\citenamefont
  {Colombo}, \citenamefont {Salerno}, \citenamefont {Gruber}, \citenamefont
  {Freppaz}, \citenamefont {Williams}, \citenamefont {Fratianni},\ and\
  \citenamefont {Giardino}}]{colombo2018}%
  \BibitemOpen
  \bibfield  {author} {\bibinfo {author} {\bibfnamefont {Nicola}\ \bibnamefont
  {Colombo}}, \bibinfo {author} {\bibfnamefont {Franco}\ \bibnamefont
  {Salerno}}, \bibinfo {author} {\bibfnamefont {Stephan}\ \bibnamefont
  {Gruber}}, \bibinfo {author} {\bibfnamefont {Michele}\ \bibnamefont
  {Freppaz}}, \bibinfo {author} {\bibfnamefont {Mark}\ \bibnamefont
  {Williams}}, \bibinfo {author} {\bibfnamefont {Simona}\ \bibnamefont
  {Fratianni}}, \ and\ \bibinfo {author} {\bibfnamefont {Marco}\ \bibnamefont
  {Giardino}},\ }\bibfield  {title} {\enquote {\bibinfo {title} {Impacts of
  permafrost degradation on inorganic chemistry of surface fresh water},}\
  }\href@noop {} {\bibfield  {journal} {\bibinfo  {journal} {Glob. Plan.
  Change}\ }\textbf {\bibinfo {volume} {162}},\ \bibinfo {pages} {69--83}
  (\bibinfo {year} {2018})}\BibitemShut {NoStop}%
\bibitem [{\citenamefont {Chadburn}\ \emph {et~al.}(2017)\citenamefont
  {Chadburn}, \citenamefont {Burke}, \citenamefont {Cox}, \citenamefont
  {Friedlingstein}, \citenamefont {Hugelius},\ and\ \citenamefont
  {Westermann}}]{chadburn2017}%
  \BibitemOpen
  \bibfield  {author} {\bibinfo {author} {\bibfnamefont {SE}~\bibnamefont
  {Chadburn}}, \bibinfo {author} {\bibfnamefont {EJ}~\bibnamefont {Burke}},
  \bibinfo {author} {\bibfnamefont {PM}~\bibnamefont {Cox}}, \bibinfo {author}
  {\bibfnamefont {P}~\bibnamefont {Friedlingstein}}, \bibinfo {author}
  {\bibfnamefont {Gustaf}\ \bibnamefont {Hugelius}}, \ and\ \bibinfo {author}
  {\bibfnamefont {S}~\bibnamefont {Westermann}},\ }\bibfield  {title} {\enquote
  {\bibinfo {title} {An observation-based constraint on permafrost loss as a
  function of global warming},}\ }\href@noop {} {\bibfield  {journal} {\bibinfo
   {journal} {Nat. Clim. Chang.}\ }\textbf {\bibinfo {volume} {7}},\ \bibinfo
  {pages} {340--344} (\bibinfo {year} {2017})}\BibitemShut {NoStop}%
\bibitem [{\citenamefont {Obu}\ \emph {et~al.}(2019)\citenamefont {Obu},
  \citenamefont {Westermann}, \citenamefont {Bartsch}, \citenamefont
  {Berdnikov}, \citenamefont {Christiansen}, \citenamefont {Dashtseren},
  \citenamefont {Delaloye}, \citenamefont {Elberling}, \citenamefont
  {Etzelm{\"u}ller}, \citenamefont {Kholodov} \emph {et~al.}}]{obu2019}%
  \BibitemOpen
  \bibfield  {author} {\bibinfo {author} {\bibfnamefont {Jaroslav}\
  \bibnamefont {Obu}}, \bibinfo {author} {\bibfnamefont {Sebastian}\
  \bibnamefont {Westermann}}, \bibinfo {author} {\bibfnamefont {Annett}\
  \bibnamefont {Bartsch}}, \bibinfo {author} {\bibfnamefont {Nikolai}\
  \bibnamefont {Berdnikov}}, \bibinfo {author} {\bibfnamefont {Hanne~H}\
  \bibnamefont {Christiansen}}, \bibinfo {author} {\bibfnamefont {Avirmed}\
  \bibnamefont {Dashtseren}}, \bibinfo {author} {\bibfnamefont {Reynald}\
  \bibnamefont {Delaloye}}, \bibinfo {author} {\bibfnamefont {Bo}~\bibnamefont
  {Elberling}}, \bibinfo {author} {\bibfnamefont {Bernd}\ \bibnamefont
  {Etzelm{\"u}ller}}, \bibinfo {author} {\bibfnamefont {Alexander}\
  \bibnamefont {Kholodov}},  \emph {et~al.},\ }\bibfield  {title} {\enquote
  {\bibinfo {title} {{Northern Hemisphere permafrost map based on TTOP
  modelling for 2000--2016 at 1 km$^2$ scale}},}\ }\href@noop {} {\bibfield
  {journal} {\bibinfo  {journal} {Earth-Science Rev.}\ }\textbf {\bibinfo
  {volume} {193}},\ \bibinfo {pages} {299--316} (\bibinfo {year}
  {2019})}\BibitemShut {NoStop}%
\bibitem [{\citenamefont {Dankers}\ \emph {et~al.}(2011)\citenamefont
  {Dankers}, \citenamefont {Burke},\ and\ \citenamefont {Price}}]{dankers2011}%
  \BibitemOpen
  \bibfield  {author} {\bibinfo {author} {\bibfnamefont {R}~\bibnamefont
  {Dankers}}, \bibinfo {author} {\bibfnamefont {EJ}~\bibnamefont {Burke}}, \
  and\ \bibinfo {author} {\bibfnamefont {J}~\bibnamefont {Price}},\ }\bibfield
  {title} {\enquote {\bibinfo {title} {{Simulation of permafrost and seasonal
  thaw depth in the JULES land surface scheme}},}\ }\href@noop {} {\bibfield
  {journal} {\bibinfo  {journal} {The Cryosphere}\ }\textbf {\bibinfo {volume}
  {5}},\ \bibinfo {pages} {773--790} (\bibinfo {year} {2011})}\BibitemShut
  {NoStop}%
\bibitem [{\citenamefont {Wania}\ \emph {et~al.}(2009)\citenamefont {Wania},
  \citenamefont {Ross},\ and\ \citenamefont {Prentice}}]{wania2009}%
  \BibitemOpen
  \bibfield  {author} {\bibinfo {author} {\bibfnamefont {R}~\bibnamefont
  {Wania}}, \bibinfo {author} {\bibfnamefont {I}~\bibnamefont {Ross}}, \ and\
  \bibinfo {author} {\bibfnamefont {IC}~\bibnamefont {Prentice}},\ }\bibfield
  {title} {\enquote {\bibinfo {title} {Integrating peatlands and permafrost
  into a dynamic global vegetation model: 1. evaluation and sensitivity of
  physical land surface processes},}\ }\href@noop {} {\bibfield  {journal}
  {\bibinfo  {journal} {Global Biogeochemical Cycles}\ }\textbf {\bibinfo
  {volume} {23}} (\bibinfo {year} {2009})}\BibitemShut {NoStop}%
\bibitem [{\citenamefont {Guimberteau}\ \emph {et~al.}(2018)\citenamefont
  {Guimberteau}, \citenamefont {Zhu}, \citenamefont {Maignan}, \citenamefont
  {Huang}, \citenamefont {Yue}, \citenamefont {Dantec-N{\'e}d{\'e}lec},
  \citenamefont {Ottl{\'e}}, \citenamefont {Jornet-Puig}, \citenamefont
  {Bastos}, \citenamefont {Laurent} \emph {et~al.}}]{guimberteau2018}%
  \BibitemOpen
  \bibfield  {author} {\bibinfo {author} {\bibfnamefont {Matthieu}\
  \bibnamefont {Guimberteau}}, \bibinfo {author} {\bibfnamefont {Dan}\
  \bibnamefont {Zhu}}, \bibinfo {author} {\bibfnamefont {Fabienne}\
  \bibnamefont {Maignan}}, \bibinfo {author} {\bibfnamefont {Ye}~\bibnamefont
  {Huang}}, \bibinfo {author} {\bibfnamefont {Chao}\ \bibnamefont {Yue}},
  \bibinfo {author} {\bibfnamefont {Sarah}\ \bibnamefont
  {Dantec-N{\'e}d{\'e}lec}}, \bibinfo {author} {\bibfnamefont {Catherine}\
  \bibnamefont {Ottl{\'e}}}, \bibinfo {author} {\bibfnamefont {Albert}\
  \bibnamefont {Jornet-Puig}}, \bibinfo {author} {\bibfnamefont {Ana}\
  \bibnamefont {Bastos}}, \bibinfo {author} {\bibfnamefont {Pierre}\
  \bibnamefont {Laurent}},  \emph {et~al.},\ }\bibfield  {title} {\enquote
  {\bibinfo {title} {{ORCHIDEE-MICT (v8. 4.1), a land surface model for the
  high latitudes: model description and validation}},}\ }\href@noop {}
  {\bibfield  {journal} {\bibinfo  {journal} {Geosci. Model Dev.}\ }\textbf
  {\bibinfo {volume} {11}},\ \bibinfo {pages} {121--163} (\bibinfo {year}
  {2018})}\BibitemShut {NoStop}%
\bibitem [{\citenamefont {Lawrence}\ \emph {et~al.}(2008)\citenamefont
  {Lawrence}, \citenamefont {Slater}, \citenamefont {Romanovsky},\ and\
  \citenamefont {Nicolsky}}]{lawrence2008}%
  \BibitemOpen
  \bibfield  {author} {\bibinfo {author} {\bibfnamefont {David~M}\ \bibnamefont
  {Lawrence}}, \bibinfo {author} {\bibfnamefont {Andrew~G}\ \bibnamefont
  {Slater}}, \bibinfo {author} {\bibfnamefont {Vladimir~E}\ \bibnamefont
  {Romanovsky}}, \ and\ \bibinfo {author} {\bibfnamefont {Dmitry~J}\
  \bibnamefont {Nicolsky}},\ }\bibfield  {title} {\enquote {\bibinfo {title}
  {Sensitivity of a model projection of near-surface permafrost degradation to
  soil column depth and representation of soil organic matter},}\ }\href@noop
  {} {\bibfield  {journal} {\bibinfo  {journal} {J. Geophys. Res.: Earth
  Surface}\ }\textbf {\bibinfo {volume} {113}} (\bibinfo {year}
  {2008})}\BibitemShut {NoStop}%
\bibitem [{\citenamefont {Andresen}\ \emph {et~al.}(2020)\citenamefont
  {Andresen}, \citenamefont {Lawrence}, \citenamefont {Wilson}, \citenamefont
  {McGuire}, \citenamefont {Koven}, \citenamefont {Schaefer}, \citenamefont
  {Jafarov}, \citenamefont {Peng}, \citenamefont {Chen}, \citenamefont
  {Gouttevin} \emph {et~al.}}]{andresen2020}%
  \BibitemOpen
  \bibfield  {author} {\bibinfo {author} {\bibfnamefont {Christian~G}\
  \bibnamefont {Andresen}}, \bibinfo {author} {\bibfnamefont {David~M}\
  \bibnamefont {Lawrence}}, \bibinfo {author} {\bibfnamefont {Cathy~J}\
  \bibnamefont {Wilson}}, \bibinfo {author} {\bibfnamefont {A~David}\
  \bibnamefont {McGuire}}, \bibinfo {author} {\bibfnamefont {Charles}\
  \bibnamefont {Koven}}, \bibinfo {author} {\bibfnamefont {Kevin}\ \bibnamefont
  {Schaefer}}, \bibinfo {author} {\bibfnamefont {Elchin}\ \bibnamefont
  {Jafarov}}, \bibinfo {author} {\bibfnamefont {Shushi}\ \bibnamefont {Peng}},
  \bibinfo {author} {\bibfnamefont {Xiaodong}\ \bibnamefont {Chen}}, \bibinfo
  {author} {\bibfnamefont {Isabelle}\ \bibnamefont {Gouttevin}},  \emph
  {et~al.},\ }\bibfield  {title} {\enquote {\bibinfo {title} {{Soil moisture
  and hydrology projections of the permafrost region - a model
  intercomparison}},}\ }\href@noop {} {\bibfield  {journal} {\bibinfo
  {journal} {The Cryosphere}\ }\textbf {\bibinfo {volume} {14}},\ \bibinfo
  {pages} {445--459} (\bibinfo {year} {2020})}\BibitemShut {NoStop}%
\bibitem [{\citenamefont {Beckermann}\ \emph {et~al.}(1999)\citenamefont
  {Beckermann}, \citenamefont {Diepers}, \citenamefont {Steinbach},
  \citenamefont {Karma},\ and\ \citenamefont {Tong}}]{beckermann1999modeling}%
  \BibitemOpen
  \bibfield  {author} {\bibinfo {author} {\bibfnamefont {Christoph}\
  \bibnamefont {Beckermann}}, \bibinfo {author} {\bibfnamefont {H-J}\
  \bibnamefont {Diepers}}, \bibinfo {author} {\bibfnamefont {Ingo}\
  \bibnamefont {Steinbach}}, \bibinfo {author} {\bibfnamefont {Alain}\
  \bibnamefont {Karma}}, \ and\ \bibinfo {author} {\bibfnamefont {Xinglin}\
  \bibnamefont {Tong}},\ }\bibfield  {title} {\enquote {\bibinfo {title}
  {Modeling melt convection in phase-field simulations of solidification},}\
  }\href@noop {} {\bibfield  {journal} {\bibinfo  {journal} {J. Comput. Phys.}\
  }\textbf {\bibinfo {volume} {154}},\ \bibinfo {pages} {468--496} (\bibinfo
  {year} {1999})}\BibitemShut {NoStop}%
\bibitem [{\citenamefont {Favier}\ \emph {et~al.}(2019)\citenamefont {Favier},
  \citenamefont {Purseed},\ and\ \citenamefont
  {Duchemin}}]{favier2019rayleigh}%
  \BibitemOpen
  \bibfield  {author} {\bibinfo {author} {\bibfnamefont {Benjamin}\
  \bibnamefont {Favier}}, \bibinfo {author} {\bibfnamefont {Jhaswantsing}\
  \bibnamefont {Purseed}}, \ and\ \bibinfo {author} {\bibfnamefont {Laurent}\
  \bibnamefont {Duchemin}},\ }\bibfield  {title} {\enquote {\bibinfo {title}
  {{Rayleigh--B{\'e}nard convection with a melting boundary}},}\ }\href@noop {}
  {\bibfield  {journal} {\bibinfo  {journal} {J. Fluid Mech.}\ }\textbf
  {\bibinfo {volume} {858}},\ \bibinfo {pages} {437--473} (\bibinfo {year}
  {2019})}\BibitemShut {NoStop}%
\bibitem [{\citenamefont {Yang}\ \emph {et~al.}(2022)\citenamefont {Yang},
  \citenamefont {Chong}, \citenamefont {Liu}, \citenamefont {Verzicco},\ and\
  \citenamefont {Lohse}}]{yang2022abrupt}%
  \BibitemOpen
  \bibfield  {author} {\bibinfo {author} {\bibfnamefont {Rui}\ \bibnamefont
  {Yang}}, \bibinfo {author} {\bibfnamefont {Kai~Leong}\ \bibnamefont {Chong}},
  \bibinfo {author} {\bibfnamefont {Hao-Ran}\ \bibnamefont {Liu}}, \bibinfo
  {author} {\bibfnamefont {Roberto}\ \bibnamefont {Verzicco}}, \ and\ \bibinfo
  {author} {\bibfnamefont {Detlef}\ \bibnamefont {Lohse}},\ }\bibfield  {title}
  {\enquote {\bibinfo {title} {Abrupt transition from slow to fast melting of
  ice},}\ }\href@noop {} {\bibfield  {journal} {\bibinfo  {journal} {Phys. Rev.
  Fluids}\ }\textbf {\bibinfo {volume} {7}},\ \bibinfo {pages} {083503}
  (\bibinfo {year} {2022})}\BibitemShut {NoStop}%
\bibitem [{\citenamefont {Hester}\ \emph {et~al.}(2020)\citenamefont {Hester},
  \citenamefont {Couston}, \citenamefont {Favier}, \citenamefont {Burns},\ and\
  \citenamefont {Vasil}}]{hester2020improved}%
  \BibitemOpen
  \bibfield  {author} {\bibinfo {author} {\bibfnamefont {Eric~W}\ \bibnamefont
  {Hester}}, \bibinfo {author} {\bibfnamefont {Louis-Alexandre}\ \bibnamefont
  {Couston}}, \bibinfo {author} {\bibfnamefont {Benjamin}\ \bibnamefont
  {Favier}}, \bibinfo {author} {\bibfnamefont {Keaton~J}\ \bibnamefont
  {Burns}}, \ and\ \bibinfo {author} {\bibfnamefont {Geoffrey~M}\ \bibnamefont
  {Vasil}},\ }\bibfield  {title} {\enquote {\bibinfo {title} {Improved
  phase-field models of melting and dissolution in multi-component flows},}\
  }\href@noop {} {\bibfield  {journal} {\bibinfo  {journal} {Proc. Royal Soc.
  A}\ }\textbf {\bibinfo {volume} {476}},\ \bibinfo {pages} {20200508}
  (\bibinfo {year} {2020})}\BibitemShut {NoStop}%
\bibitem [{\citenamefont {Rees}\ and\ \citenamefont
  {Pop}(2000)}]{rees2000vertical}%
  \BibitemOpen
  \bibfield  {author} {\bibinfo {author} {\bibfnamefont {Andrew}\ \bibnamefont
  {Rees}}\ and\ \bibinfo {author} {\bibfnamefont {Ioan}\ \bibnamefont {Pop}},\
  }\bibfield  {title} {\enquote {\bibinfo {title} {Vertical free convective
  boundary-layer flow in a porous medium using a thermal nonequilibrium
  model},}\ }\href@noop {} {\bibfield  {journal} {\bibinfo  {journal} {J.
  Porous Media}\ }\textbf {\bibinfo {volume} {3}},\ \bibinfo {pages} {31}
  (\bibinfo {year} {2000})}\BibitemShut {NoStop}%
\bibitem [{\citenamefont {Wang}\ \emph {et~al.}(2021)\citenamefont {Wang},
  \citenamefont {Calzavarini}, \citenamefont {Sun},\ and\ \citenamefont
  {Toschi}}]{wang2021growth}%
  \BibitemOpen
  \bibfield  {author} {\bibinfo {author} {\bibfnamefont {Ziqi}\ \bibnamefont
  {Wang}}, \bibinfo {author} {\bibfnamefont {Enrico}\ \bibnamefont
  {Calzavarini}}, \bibinfo {author} {\bibfnamefont {Chao}\ \bibnamefont {Sun}},
  \ and\ \bibinfo {author} {\bibfnamefont {Federico}\ \bibnamefont {Toschi}},\
  }\bibfield  {title} {\enquote {\bibinfo {title} {How the growth of ice
  depends on the fluid dynamics underneath},}\ }\href@noop {} {\bibfield
  {journal} {\bibinfo  {journal} {Proc. Natl. Acad. Sciences}\ }\textbf
  {\bibinfo {volume} {118}},\ \bibinfo {pages} {e2012870118} (\bibinfo {year}
  {2021})}\BibitemShut {NoStop}%
\bibitem [{\citenamefont {Lapwood}(1948)}]{lapwood1948convection}%
  \BibitemOpen
  \bibfield  {author} {\bibinfo {author} {\bibfnamefont {E.~R.}\ \bibnamefont
  {Lapwood}},\ }\bibfield  {title} {\enquote {\bibinfo {title} {Convection of a
  fluid in a porous medium},}\ }\href@noop {} {\bibfield  {journal} {\bibinfo
  {journal} {Math. Proc. Cambridge Phil. Soc.}\ }\textbf {\bibinfo {volume}
  {44}},\ \bibinfo {pages} {508} (\bibinfo {year} {1948})}\BibitemShut
  {NoStop}%
\bibitem [{\citenamefont {Banu}\ and\ \citenamefont
  {Rees}(2002)}]{banu2002onset}%
  \BibitemOpen
  \bibfield  {author} {\bibinfo {author} {\bibfnamefont {Nurzahan}\
  \bibnamefont {Banu}}\ and\ \bibinfo {author} {\bibfnamefont {DAS}\
  \bibnamefont {Rees}},\ }\bibfield  {title} {\enquote {\bibinfo {title} {Onset
  of darcy--benard convection using a thermal non-equilibrium model},}\
  }\href@noop {} {\bibfield  {journal} {\bibinfo  {journal} {Int. J. Heat Mass
  Trans.}\ }\textbf {\bibinfo {volume} {45}},\ \bibinfo {pages} {2221--2228}
  (\bibinfo {year} {2002})}\BibitemShut {NoStop}%
\bibitem [{\citenamefont {Nield}\ \emph {et~al.}(2006)\citenamefont {Nield},
  \citenamefont {Bejan} \emph {et~al.}}]{nield2006convection}%
  \BibitemOpen
  \bibfield  {author} {\bibinfo {author} {\bibfnamefont {Donald~A}\
  \bibnamefont {Nield}}, \bibinfo {author} {\bibfnamefont {Adrian}\
  \bibnamefont {Bejan}},  \emph {et~al.},\ }\href@noop {} {\emph {\bibinfo
  {title} {Convection in porous media}}},\ Vol.~\bibinfo {volume} {3}\
  (\bibinfo  {publisher} {Springer},\ \bibinfo {year} {2006})\BibitemShut
  {NoStop}%
\bibitem [{\citenamefont {Rubenstein}(1971)}]{rubinshteuin1971stefan}%
  \BibitemOpen
  \bibfield  {author} {\bibinfo {author} {\bibfnamefont {L.I.}\ \bibnamefont
  {Rubenstein}},\ }\href@noop {} {\emph {\bibinfo {title} {{The Stefan
  problem}}}},\ Vol.~\bibinfo {volume} {27}\ (\bibinfo  {publisher} {American
  Mathematical Soc.},\ \bibinfo {year} {1971})\BibitemShut {NoStop}%
\bibitem [{\citenamefont {Hewitt}(2020)}]{hewitt2020}%
  \BibitemOpen
  \bibfield  {author} {\bibinfo {author} {\bibfnamefont {DR}~\bibnamefont
  {Hewitt}},\ }\bibfield  {title} {\enquote {\bibinfo {title} {Vigorous
  convection in porous media},}\ }\href@noop {} {\bibfield  {journal} {\bibinfo
   {journal} {Proc. Royal Soc. A}\ }\textbf {\bibinfo {volume} {476}},\
  \bibinfo {pages} {20200111} (\bibinfo {year} {2020})}\BibitemShut {NoStop}%
\bibitem [{\citenamefont {Boffetta}\ \emph {et~al.}(2020)\citenamefont
  {Boffetta}, \citenamefont {Borgnino},\ and\ \citenamefont
  {Musacchio}}]{boffetta2020}%
  \BibitemOpen
  \bibfield  {author} {\bibinfo {author} {\bibfnamefont {G}~\bibnamefont
  {Boffetta}}, \bibinfo {author} {\bibfnamefont {M}~\bibnamefont {Borgnino}}, \
  and\ \bibinfo {author} {\bibfnamefont {S}~\bibnamefont {Musacchio}},\
  }\bibfield  {title} {\enquote {\bibinfo {title} {{Scaling of Rayleigh-Taylor
  mixing in porous media}},}\ }\href@noop {} {\bibfield  {journal} {\bibinfo
  {journal} {Phys. Rev. Fluids}\ }\textbf {\bibinfo {volume} {5}},\ \bibinfo
  {pages} {062501} (\bibinfo {year} {2020})}\BibitemShut {NoStop}%
\bibitem [{\citenamefont {Chaigne}\ \emph {et~al.}(2023)\citenamefont
  {Chaigne}, \citenamefont {Carpy}, \citenamefont {Mass{\'e}}, \citenamefont
  {Derr}, \citenamefont {Courrech~du Pont},\ and\ \citenamefont
  {Berhanu}}]{chaigne2023emergence}%
  \BibitemOpen
  \bibfield  {author} {\bibinfo {author} {\bibfnamefont {Martin}\ \bibnamefont
  {Chaigne}}, \bibinfo {author} {\bibfnamefont {Sabrina}\ \bibnamefont
  {Carpy}}, \bibinfo {author} {\bibfnamefont {Marion}\ \bibnamefont
  {Mass{\'e}}}, \bibinfo {author} {\bibfnamefont {Julien}\ \bibnamefont
  {Derr}}, \bibinfo {author} {\bibfnamefont {Sylvain}\ \bibnamefont
  {Courrech~du Pont}}, \ and\ \bibinfo {author} {\bibfnamefont {Michael}\
  \bibnamefont {Berhanu}},\ }\bibfield  {title} {\enquote {\bibinfo {title}
  {Emergence of tip singularities in dissolution patterns},}\ }\href@noop {}
  {\bibfield  {journal} {\bibinfo  {journal} {Proc. Natl. Acad. Sciences}\
  }\textbf {\bibinfo {volume} {120}},\ \bibinfo {pages} {e2309379120} (\bibinfo
  {year} {2023})}\BibitemShut {NoStop}%
\bibitem [{\citenamefont {Boike}\ \emph {et~al.}(1998)\citenamefont {Boike},
  \citenamefont {Roth},\ and\ \citenamefont {Overduin}}]{boike1998}%
  \BibitemOpen
  \bibfield  {author} {\bibinfo {author} {\bibfnamefont {Julia}\ \bibnamefont
  {Boike}}, \bibinfo {author} {\bibfnamefont {Kurt}\ \bibnamefont {Roth}}, \
  and\ \bibinfo {author} {\bibfnamefont {Pier~Paul}\ \bibnamefont {Overduin}},\
  }\bibfield  {title} {\enquote {\bibinfo {title} {{Thermal and hydrologic
  dynamics of the active layer at a continuous permafrost site (Taymyr
  Peninsula, Siberia)}},}\ }\href@noop {} {\bibfield  {journal} {\bibinfo
  {journal} {Water Resour. Res.}\ }\textbf {\bibinfo {volume} {34}},\ \bibinfo
  {pages} {355--363} (\bibinfo {year} {1998})}\BibitemShut {NoStop}%
\bibitem [{\citenamefont {Weism{\"u}ller}\ \emph {et~al.}(2011)\citenamefont
  {Weism{\"u}ller}, \citenamefont {Wollschl{\"a}ger}, \citenamefont {Boike},
  \citenamefont {Pan}, \citenamefont {Yu},\ and\ \citenamefont
  {Roth}}]{weismuller2011}%
  \BibitemOpen
  \bibfield  {author} {\bibinfo {author} {\bibfnamefont {Jens}\ \bibnamefont
  {Weism{\"u}ller}}, \bibinfo {author} {\bibfnamefont {Ute}\ \bibnamefont
  {Wollschl{\"a}ger}}, \bibinfo {author} {\bibfnamefont {Julia}\ \bibnamefont
  {Boike}}, \bibinfo {author} {\bibfnamefont {Xicai}\ \bibnamefont {Pan}},
  \bibinfo {author} {\bibfnamefont {Q}~\bibnamefont {Yu}}, \ and\ \bibinfo
  {author} {\bibfnamefont {Kurt}\ \bibnamefont {Roth}},\ }\bibfield  {title}
  {\enquote {\bibinfo {title} {Modeling the thermal dynamics of the active
  layer at two contrasting permafrost sites on svalbard and on the tibetan
  plateau},}\ }\href@noop {} {\bibfield  {journal} {\bibinfo  {journal} {The
  Cryosphere}\ }\textbf {\bibinfo {volume} {5}},\ \bibinfo {pages} {741--757}
  (\bibinfo {year} {2011})}\BibitemShut {NoStop}%
\bibitem [{\citenamefont {Hollesen}\ \emph {et~al.}(2011)\citenamefont
  {Hollesen}, \citenamefont {Elberling},\ and\ \citenamefont
  {Jansson}}]{hollesen2011}%
  \BibitemOpen
  \bibfield  {author} {\bibinfo {author} {\bibfnamefont {J{\o}rgen}\
  \bibnamefont {Hollesen}}, \bibinfo {author} {\bibfnamefont {Bo}~\bibnamefont
  {Elberling}}, \ and\ \bibinfo {author} {\bibfnamefont {Per-Erik}\
  \bibnamefont {Jansson}},\ }\bibfield  {title} {\enquote {\bibinfo {title}
  {{Future active layer dynamics and carbon dioxide production from thawing
  permafrost layers in Northeast Greenland}},}\ }\href@noop {} {\bibfield
  {journal} {\bibinfo  {journal} {Glob. Change Biol.}\ }\textbf {\bibinfo
  {volume} {17}},\ \bibinfo {pages} {911--926} (\bibinfo {year}
  {2011})}\BibitemShut {NoStop}%
\bibitem [{\citenamefont {Gleeson}\ \emph {et~al.}(2011)\citenamefont
  {Gleeson}, \citenamefont {Smith}, \citenamefont {Moosdorf}, \citenamefont
  {Hartmann}, \citenamefont {D{\"u}rr}, \citenamefont {Manning}, \citenamefont
  {van Beek},\ and\ \citenamefont {Jellinek}}]{gleeson2011}%
  \BibitemOpen
  \bibfield  {author} {\bibinfo {author} {\bibfnamefont {Tom}\ \bibnamefont
  {Gleeson}}, \bibinfo {author} {\bibfnamefont {Leslie}\ \bibnamefont {Smith}},
  \bibinfo {author} {\bibfnamefont {Nils}\ \bibnamefont {Moosdorf}}, \bibinfo
  {author} {\bibfnamefont {Jens}\ \bibnamefont {Hartmann}}, \bibinfo {author}
  {\bibfnamefont {Hans~H}\ \bibnamefont {D{\"u}rr}}, \bibinfo {author}
  {\bibfnamefont {Andrew~H}\ \bibnamefont {Manning}}, \bibinfo {author}
  {\bibfnamefont {Ludovicus~PH}\ \bibnamefont {van Beek}}, \ and\ \bibinfo
  {author} {\bibfnamefont {A~Mark}\ \bibnamefont {Jellinek}},\ }\bibfield
  {title} {\enquote {\bibinfo {title} {{Mapping permeability over the surface
  of the Earth}},}\ }\href@noop {} {\bibfield  {journal} {\bibinfo  {journal}
  {Geophys. Res. Lett.}\ }\textbf {\bibinfo {volume} {38}} (\bibinfo {year}
  {2011})}\BibitemShut {NoStop}%
\bibitem [{\citenamefont {Chapuis}(2012)}]{chapuis2012}%
  \BibitemOpen
  \bibfield  {author} {\bibinfo {author} {\bibfnamefont {Robert~P}\
  \bibnamefont {Chapuis}},\ }\bibfield  {title} {\enquote {\bibinfo {title}
  {Predicting the saturated hydraulic conductivity of soils: a review},}\
  }\href@noop {} {\bibfield  {journal} {\bibinfo  {journal} {Bull. Eng. Geol.
  Environ.}\ }\textbf {\bibinfo {volume} {71}},\ \bibinfo {pages} {401--434}
  (\bibinfo {year} {2012})}\BibitemShut {NoStop}%
\bibitem [{\citenamefont {Ebel}\ \emph {et~al.}(2019)\citenamefont {Ebel},
  \citenamefont {Koch},\ and\ \citenamefont {Walvoord}}]{ebel2019}%
  \BibitemOpen
  \bibfield  {author} {\bibinfo {author} {\bibfnamefont {Brian~A}\ \bibnamefont
  {Ebel}}, \bibinfo {author} {\bibfnamefont {Joshua~C}\ \bibnamefont {Koch}}, \
  and\ \bibinfo {author} {\bibfnamefont {Michelle~A}\ \bibnamefont
  {Walvoord}},\ }\bibfield  {title} {\enquote {\bibinfo {title} {{Soil
  physical, hydraulic, and thermal properties in interior Alaska, USA:
  Implications for hydrologic response to thawing permafrost conditions}},}\
  }\href@noop {} {\bibfield  {journal} {\bibinfo  {journal} {Water Resour.
  Res.}\ }\textbf {\bibinfo {volume} {55}},\ \bibinfo {pages} {4427--4447}
  (\bibinfo {year} {2019})}\BibitemShut {NoStop}%
\bibitem [{\citenamefont {Zipper}\ \emph {et~al.}(2018)\citenamefont {Zipper},
  \citenamefont {Lamontagne-Hall{\'e}}, \citenamefont {McKenzie},\ and\
  \citenamefont {Rocha}}]{zipper2018}%
  \BibitemOpen
  \bibfield  {author} {\bibinfo {author} {\bibfnamefont {Samuel~C}\
  \bibnamefont {Zipper}}, \bibinfo {author} {\bibfnamefont {Pierrick}\
  \bibnamefont {Lamontagne-Hall{\'e}}}, \bibinfo {author} {\bibfnamefont
  {Jeffrey~M}\ \bibnamefont {McKenzie}}, \ and\ \bibinfo {author}
  {\bibfnamefont {Adrian~V}\ \bibnamefont {Rocha}},\ }\bibfield  {title}
  {\enquote {\bibinfo {title} {Groundwater controls on postfire permafrost
  thaw: Water and energy balance effects},}\ }\href@noop {} {\bibfield
  {journal} {\bibinfo  {journal} {J. Geophys. Res.: Earth Surface}\ }\textbf
  {\bibinfo {volume} {123}},\ \bibinfo {pages} {2677--2694} (\bibinfo {year}
  {2018})}\BibitemShut {NoStop}%
\end{thebibliography}%

\end{document}